\documentclass{article}
\usepackage{amssymb}
\usepackage{amsmath}
\usepackage{graphicx}
\usepackage{caption}
\usepackage{float}
\usepackage{subcaption}
\usepackage{algorithm}
\usepackage{algpseudocode}
\usepackage{amsfonts}
\usepackage{framed}
\usepackage[final]{corl_2025} 
\newcommand{\sect}[1]{Section~\ref{#1}}

\title{Off Policy Lyapunov Stability in Reinforcement Learning}

\author{
  Sarvan Gill\\
  Department of Mechanical Engineering\\
  University of Victoria, 
  Canada\\
  \texttt{sarvan13@uvic.ca} \\
  \And 
  Daniela Constantiescu\\
  Department of Mechanical Engineering\\
  University of Victoria, 
  Canada\\
  \texttt{danielac@uvic.ca} \\
}

\begin{document}
\maketitle


\begin{abstract}
Traditional reinforcement learning lacks the ability to provide stability guarantees. More recent algorithms learn Lyapunov functions alongside the control policies to ensure stable learning. However, the current self-learned Lyapunov functions are sample inefficient due to their on-policy nature. This paper introduces a method for learning Lyapunov functions off-policy and incorporates the proposed off-policy Lyapunov function into the Soft Actor Critic and Proximal Policy Optimization algorithms to provide them with a data efficient stability certificate. Simulations of an inverted pendulum and a quadrotor illustrate the improved performance of the two algorithms when endowed with the proposed off-policy Lyapunov function.
\end{abstract}



\section{Introduction}

Deep Reinforcement Learning (DRL) is emerging as a common robot control strategy because of its many recent promising results in challenging control tasks for systems with strongly non-linear dynamics and high dimensional state spaces, where classical control methods may struggle~\citep{drl-robot-2025}. Learning from experience is a pillar of Reinforcement Learning (RL) and an agent's success is directly tied to the experience it learns from~\citep{Sutton1998}. Given that it can be unsafe for a robot to collect trial and error samples of experience in the real world, safety and sample efficiency are important considerations for RL in robotics.

Stability is prerequisite for the safety of controlled systems. Given that unstable systems are unpredictable and can be dangerous, practical applications that require reliable and safe robots demand that the robots be guaranteed stable during task execution. Early sample-based RL techniques cannot certify stability. More recent RL methods aim to incorporate Lyapunov stability mechanisms into robot learning~\citep{robot-safety}, generally by computing a Lyapunov function for the task error of the robot in closed-loop with the RL agent. While the existence of a Lyapunov function is sufficient for stable learning, a fundamental challenge to the stability analysis of RL for robotics arises from the fact that no systematic approach exists for determining Lyapunov functions for non-linear systems. RL research has tackled this challenge for some time~\citep{barto-ly}. 

In model-based RL, Lyapunov functions that use a model of the system dynamics guarantee stability directly~\citep{berkenkamp}, while control barrier functions ensure it through certifying safety~\citep{blac, tokens}. For control affine systems with known dynamics, solving for a control Lyapunov function leads to a list of permissible stabilizing controls~\citep{ly-survey, drive-ly}. In model-free RL, a backup safe controller can be included to guarantee stability~\cite{chen19}, including during online training~\citep{calf}. Otherwise, model-free RL must turn to sample-based stability guarantees, generally by starting with a candidate Lyapunov function and then finding a control policy that makes the candidate Lyapunov~\citep{chow-ly, lac, LBAC}.  In this approach, the value function often serves as the candidate and the reward function must be reshaped into a cost whose minimum has a value of zero at the equilibrium of the system. Alternatively, better performance can be achieved by learning a neural Lyapunov function through a Lyapunov risk loss function which penalizes the neural network for any violations of the Lyapunov conditions~\cite{almost-ly}. A self-learned neural Lyapunov function neither requires changes to the reward function nor restricts the candidate to being the RL value function.

Whereas a candidate Lyapunov function has the advantage of being able to use off-policy data to increase sample efficiency during training, learning a neural Lyapunov function directly dispenses with the overhead in creating a cost function and can produce better results \citep{almost-ly, twin-ly}. However, to the authors' best knowledge, direct learning of a neural Lyapunov function is currently limited to learning off on-policy data. This paper introduces a novel method to learn a Lyapunov function directly using either on-policy or off-policy data. The proposed method extends~\citep{almost-ly} to account for off-policy data and includes an additional hyper parameter that controls a minimum rate of decay of the Lyapunov function. In essence, the paper:
\begin{itemize}
    \item provides a framework that can learn Lyapunov functions off policy;
    \item demonstrates how the off-policy Lyapunov functions can guide state-of-the-art RL algorithms to learn stable policies;
    \item illustrates that, compared to other stable RL methods, the proposed framework can increase sample efficiency without performance sacrifices.
\end{itemize}

\section{Preliminaries}

This section briefly recalls the definitions needed to introduce the proposed off-policy Lyapunov functions in \sect{sec:off-policy Lyapunov}.

Consider the closed-loop system 

\begin{equation}
    \dot{\mathbf{x}}(t) = f (\mathbf{x}(t), u(\mathbf{x}(t))), \mathbf{x}_0 = \mathbf{x}(0)\textrm{,}
    \label{eq:dynamics}
\end{equation}

with state $\mathbf{x}(t) \in \mathcal{X} \subseteq \mathbb{R}^n$, control signal $u(\mathbf{x}(t)): \mathcal{X} \rightarrow \mathbb{R}^m$, and continuous non-linear dynamics $f:\mathcal{X} \rightarrow \mathbb{R}^n$. 

\textbf{Lyapunov Stability}

An equilibrium state $\mathbf{x}_e \in \mathcal{X}$ of the closed-loop system~(\ref{eq:dynamics}) is Lyapunov stable if for every $\epsilon \in \mathbb{R}_{> 0}$ there exists $\delta \in \mathbb{R}_{>0}$ such that $||\mathbf{x}_0 - \mathbf{x}_e|| < \delta$ implies $||\mathbf{x(t)} - \mathbf{x}_e|| < \epsilon$ for all $t > 0$. The equilibrium $\mathbf{x}_e$ is asymptotically stable if it is Lyapunov stable and there exists $\delta \in \mathbb{R}_{>0}$ such that $||\mathbf{x}_0 - \mathbf{x}_e|| < \delta$ implies $\lim_{t \rightarrow \infty} \mathbf{x(t)} = \mathbf{x_e}$.

\textbf{Lyapunov Stability Criterion}

An equilibrium state $\mathbf{x}_e \in \mathcal{X}$ of the closed-loop system~(\ref{eq:dynamics}) is Lyapunov stable if the system admits a Lyapunov function, that is, a positive semi-definite function $L: \mathcal{X} \rightarrow \mathbb{R}_{\ge 0}$ whose value is zero at the equilibrium, $L(\mathbf{x}_e)=0$, and is positive at every other state, $L(\mathbf{x})>0$ for all $\mathbf{x}\ne \mathbf{x}_e$, and whose Lie derivative is non-positive along all system trajectories, $\mathcal{L}_f L(\mathbf{x})=\nabla L\cdot  f (\mathbf{x}(t), u(\mathbf{x}(t)))\le 0$. The equilibrium $\mathbf{x}_e$ is asymptotically stable if the Lie derivative of the Lyapunov function is strictly negative, $\mathcal{L}_f L(\mathbf{x}) < 0$.

\textbf{Neural Lyapunov Functions}

While the existence of a Lyapunov function $L$ is sufficient to certify the stability of the equilibrium $\mathbf{x}_e$, classical control theory offers no analytical method for deriving such a Lyapunov function. However, recent work~\citep{nlpc, physics-nlp} has shown that parametrized neural networks can estimate Lyapunov functions. For control tasks, effective and certifiable Lyapunov functions can be learned by training a neural Lyapunov function $L_\theta$ to minimize the Lyapunov risk $J_{L_\theta} $ over an on-policy dataset $\mathcal{B}$~\citep{nlpc}:

\begin{equation}
    J_{L_\theta} = \mathbb{E}_{\mathcal{B}} \left [ \max(0, -L_\theta(\mathbf{x})) + \max(0, \mathcal{L}_f L_\theta(\mathbf{x})) + L_\theta(\mathbf{x}_e)^2 \right] 
    \label{eq:ly-risk}
\end{equation}

\textbf{Reinforcement Learning}

This paper considers a dynamical system that can be modeled by a Markov Decision process (MDP). Namely, the system is defined by the interaction of an RL agent with an environment. As the result of an action $a_t \in \mathcal{A} \subset{\mathbb{R}^m}$ taken by the agent at time $t$, the state $s_t \in \mathcal{S} \subset{\mathbb{R}^n}$ of the system changes to a new state $s_{t+1}$ with probability $P(s_{t+1}|s_t, a_t)$. These transition probabilities define the system dynamics. Upon associating a reward function $R(s_t,a_t,s_{t+1})$ with the transition from $s_t$ to $s_{t+1}$ under $a_t$, the RL agent aims to learn a policy $\pi(a_t|s_t)$ that maximizes the reward it receives, typically parameterized as a neural network. In model-free RL, the transition probabilities and the reward function are not visible to the agent. Instead, the environment provides the appropriate signals, i.e., $r_t$ and $s_{t+1}$ are provided to the agent after taking an action $a_t$ in state $s_t$. The RL agent seeks to maximize the total expected return, $J = \mathbb{E}_\pi \left [ \sum_{t=0}^\infty \gamma^t r_t \right]$, where $\gamma$ is a discount factor that weights the value of future returns. This paper considers robotic systems in closed loop with an RL agent whose goal is to drive the robot to a goal state $s_G \in \mathcal{S}$. 

\textbf{Lyapunov Control in Reinforcement Learning}

For an MDP, the Lie derivative of the Lyapunov function, $\mathcal{L}_f L_\theta$, can be modeled using the following finite difference derivative~\citep{almost-ly}:

\begin{equation}
    \mathcal{L}_{f, \Delta t} L = \frac{L(s') - L(s)} {\Delta t}
    \textrm{,}
    \label{eq:op-lie}
\end{equation}

where $s$ and $s'$ are the two consecutive states with time difference $\Delta t$. When the data is sampled from the same policy, the finite difference of the Lyapunov function approximates its Lie derivative effectively. For on-policy data, the RL agent can self-learn Lyapunov functions by replacing the Lie derivative $\mathcal{L}_f L_\theta (\mathbf{x)}$ in \eqref{eq:ly-risk} with $\mathcal{L}_{f, \Delta t} L$ in \eqref{eq:op-lie}, resulting in the following Lyapunov risk~\citep{almost-ly}:

\begin{equation}
    J_{L_\theta} = \mathbb{E}_{(s, a, r, s') \sim \mathcal{B}} \left [ \max(0, -L_\theta(s) + \max(0, \mathcal{L}_{f, \Delta t} L_\theta) + L_\theta(s_G)^2 \right] 
    \label{eq:op-ly}
\end{equation}

The Lyapunov risk~\eqref{eq:op-ly} can then be used to train a Lyapunov function alongside an RL policy.

\section{Learning Off-Policy Lyapunov Functions}\label{sec:off-policy Lyapunov}

This section proposes to extend the Lyapunov risk~\eqref{eq:op-ly} to account for off-policy data. The inspiration comes from~\citep{lac}, where the RL action-value function $Q(s,a)$ serves as a Lyapunov candidate and the Lyapunov function is evaluated as the expectation over the actions under the current policy. 

Instead of using a predetermined candidate, an off-policy self-learned Lyapunov function can be determined in two steps. In a first step, similarly to~\citep{lac}, the RL agent learns a neural Lyapunov function that depends both on the state and on the action. In a second step, the agent uses the expectation over the actions under the current policy to verify the Lyapunov conditions. 

Formally, the agent learns a neural Lyapunov function $L_\eta (s,a)$ which is trained on the updated Lyapunov risk~\eqref{eq:van-lya-risk} with the redefined finite difference Lie derivative~\eqref{eq:off-lie} calculated over an off-policy dataset $D$:

\begin{equation}
    J_L(\eta) = \mathbb{E}_{(s,a,r,s') \sim D} \big[ \max(0,-L_\eta(s,a)) + \max(0, \mathcal{L}_{f, \Delta t} L_\eta) \big]  + L_\eta(s_G, \pi (s_G))^2
    \label{eq:van-lya-risk}
\end{equation}
\begin{equation}
    \mathcal{L}_{f, \Delta t} L_\eta = \frac{L_\eta(s', \pi(s')) - L_\eta(s, a)} {\Delta t}
    \label{eq:off-lie}
\end{equation}

The key differences between~\eqref{eq:van-lya-risk} and~\eqref{eq:op-ly} are the Lie derivative and the equilibrium value. The Lie derivative in~\eqref{eq:off-lie} is explicitly dependent on the current policy, as in~\citep{lac}, where the decreasing condition serves to transform the RL action-value function into a Lyapunov function. This explicit dependence on the policy is necessary for off-policy learning as the data is no longer sampled under the same policy. Intuitively, the Lie derivative in~\eqref{eq:off-lie} is now dependent on the action that the current policy would take if it ended up in some state $s'$. Furthermore, the minimum of~\eqref{eq:van-lya-risk} also depends on the action taken there under the current policy.

To verify that the function learned by the risk~\eqref{eq:van-lya-risk} is a Lyapunov function as required by the Lyapunov stability criterion, consider the expectation of $L_\eta(s,a)$:

\begin{equation}
    L_\eta(s) = \mathbb{E}_{a \sim \pi} L_\eta(s,a)
    \textrm{.}
    \label{eq:expect}
\end{equation}

Note that $L_\eta(s,a) > 0$ and $L_\eta(s_G,\pi(s_G)) = 0$ together imply that $L_\eta(s) > 0$ and $L_\eta(s_G) = 0$. Furthermore, as shown in~\citep{lac}, $\mathcal{L}_{f, \Delta t} L_\eta(s,a)) < 0$ is sufficient for the Lie derivative of $L_\eta(s)$ to decrease along any system trajectory, $\mathcal{L}_{f, \Delta t} L_\eta(s)) < 0$. 

\subsection{Practical Changes}

In practice, the RL agent learns the Lyapunov function with the help of a hyperparameter $\mu \in \mathbb{R}_{>0}$ which defines a minimum rate of decrease:

\begin{equation}
    J_L(\eta) = \mathbb{E}_{(s,a,r,s') \sim D} \big[ \max(0,-L_\eta(s,a))  + \max(0,\mathcal{L}_{f, \Delta t} L_\eta + \mu) \big] +  L_\eta(s_G, \pi (s_G))^2 
\label{eq:lya-risk}
\end{equation}

and, thus, offers the ability to scale the changes in the Lyapunov function. While the shape of the function is sufficient to guarantee stability, a degree of control over its minimum rate of decrease can be used to impact the learning of the policy and the relative weight of the Lyapunov function in the policy update. 

Imposing a minimum rate of decrease on the learned function makes it non-differentiable at the equilibrium of the system. The lack of a derivative at the equilibrium does not hinder the function from certifying stability because its Lie derivative can still be guaranteed negative everywhere but at the equilibrium. However, an important consideration is that the proposed loss function~\eqref{eq:lya-risk} cannot be zero by design, as it cannot decrease further by the required amount $\mu$ at the system equilibrium where it achieves its minimum. This issue can be side-stepped by using~\eqref{eq:lya-risk} to train the Lyapunov function and by using~\eqref{eq:van-lya-risk} to guarantee stability. Then, given the Lyapunov function learned by~\eqref{eq:lya-risk}, the system is stable if~\eqref{eq:van-lya-risk} is satisfied.

\subsection{Learning Stable Policies}

This section demonstrates how the learned off-policy Lyapunov function~\eqref{eq:lya-risk} can be used to learn stable RL policies. It builds a Lyapunov Soft Actor Critic (LSAC) algorithm by adding the off-policy Lyapunov function to guide the Soft Actor Critic Algorithm~\citep{sac} to learn the control policy. It also shows that the proposed off-policy Lyapunov function can be applied to on-policy data by building a Lyapunov Proximal Policy Optimization (LPPO) based on the Proximal Policy Optimization Algorithm~\citep{ppo}.

\textbf{Stabilizing Off-Policy Algorithms}

The SAC algorithm learns the parameterized policy via maximizing entropy using the loss function:

\begin{equation}
    J_\pi (\phi) = \mathbb{E}_{(s,a,r,s') \sim D} \big[\alpha (\log(\pi_\phi(a|s)) + \mathcal{H}) - Q_\theta(s,a) \big],
    \label{eq:sac-loss}
\end{equation}

where $\mathcal{H}$ is the minimum entropy and $\alpha$ is the entropy temperature hyperparameter which weighs the relative importance of the entropy.

The proposed LSAC first learns the off-policy Lyapunov function via~\eqref{eq:lya-risk}, and then uses it to guide the learning of the control policy through inntroducing the Lie derivative into the SAC policy loss via a Lyapunov temperature hyperparameter $\beta$ by:

\begin{equation}
    J_\pi (\phi) = \mathbb{E}_{(s,a,r,s') \sim D} \big[\alpha (\log(\pi_\phi(a|s)) + \mathcal{H}) - Q(s,a) + \beta \max(0, \mathcal{L}_{f, \Delta t} L_\eta + \mu)\big]
    \label{eq:lsac-loss}
\end{equation}

If the Lie derivative is negative by the minimum amount $\mu$, then the Lyapunov function does not bias the learning. The agent is only penalized for taking actions that cause the Lie derivative to be positive. 

Figure \ref{fig:algorithms} shows the full algorithm.

\textbf{Extension to On-Policy Algorithms}

PPO is an on-policy algorithm that learns a policy that maximizes the advantage $\hat{A}_t$, which measures the difference between the state-action pair and the expected value of the state, using the following loss function:

\begin{equation}
    J_\pi(\phi) = \mathbb{E}_{(s,a,r,s') \sim D} \left[ 
        \min \left(
            \frac{\pi_\phi}{\pi_{old}} \hat{A}_t,\ 
            \text{clip}\left(\frac{\pi_\phi}{\pi_{old}}, 1 - \epsilon, 1 + \epsilon\right) \hat{A}_t
        \right)
    \right]
    \label{eq:ppo}
\end{equation}

The hyperparameter $\epsilon$ controls the clipping of the ratio of the current policy to the sampled policy to prevent large changes in the policy. 

The proposed LPPO learns the Lyapunov function using on-policy data, similarly to POLYC~\citep{almost-ly}, but using the off-policy Lyapunov function with the loss defined in \eqref{eq:lya-risk}. Then, it includes the Lyapunov decreasing condition in an augmented advantage $\hat{A}_\beta$ by:

\begin{equation}
    \hat{A}_\beta = \hat{A}_t + \beta \min(0, -(\mathcal{L}_{f, \Delta t} L_\eta + \mu))
    \label{eq:adv}
\end{equation}

and replaces $\hat{A_t}$ with $\hat{A}_\beta$ in~\eqref{eq:ppo} in the policy loss function~\eqref{eq:ppo}:

\begin{equation}
    J_\pi(\phi) = \mathbb{E}_{(s,a,r,s') \sim D} \left[\min \left(
            \frac{\pi_\phi}{\pi_{old}} \hat{A}_\beta,\ 
            \text{clip}\left(\frac{\pi_\phi}{\pi_{old}}, 1 - \epsilon, 1 + \epsilon\right) \hat{A}_\beta
        \right)\right]
        \label{eq:lppo}
\end{equation}

As in LSAC, a negative Lie derivative does not bias the learning and a penalty is applied to the advantage when the Lie derivative is positive. 

Figure \ref{fig:algorithms} presents the full algorithm.

\textbf{Stability Certification}
Stability certificates can be obtained: (i) from the loss~\eqref{eq:lya-risk}, which indicates that the Lyapunov conditions are satisfied and a Lyapunov function is found when it converges to zero; and (ii) from the Almost Lyapunov Conditions~\citep{gao}, which certify stability when a small number of bounded violations exist near the equilibrium. The Pendulum-v1 experiment illustrates each method in Figure \ref{fig:pend_rew}~(b) and in Figure~\ref{fig:pend_almost}, respectively.

\begin{figure*}[t]
\centering
\begin{minipage}[t]{0.48\textwidth}
\raggedright
\textbf{Lyapunov Soft Actor-Critic (LSAC)}
\vspace{0.5em}
\begin{algorithmic}[1]
\State Initialize policy $\pi_\phi$, RL value function and action value function $Q_{\theta}$, $V_{\psi}$, $V_{\bar{\psi}}$, Lyapunov function $L_\eta$ randomly
\State Initialize replay buffer $\mathcal{D} \leftarrow \emptyset$
\While{steps $ < K$}
  \For{each environment step}
    \State Sample $a_t \sim \pi_\phi(a_t|s_t)$
    \State Sample $s_{t+1} \sim P(s_{t+1}|s_t, a_t)$
    \State $\mathcal{D} \leftarrow \mathcal{D} \cup \{(s_t, a_t, r_t, s_{t+1})\}$
    \State steps $\leftarrow$ steps $+ 1$
  \EndFor
  \For{each Lyapunov optimization step}
    \State Sample mini batch from $\mathcal{D}$
    \State Compute $J_{L_\eta}$ via \eqref{eq:lya-risk}
    \State $\eta \leftarrow \eta - \alpha_\eta \nabla_\eta J_{L_\eta}(\eta)$
  \EndFor
  \For{each policy optimization step}
    \State Sample mini-batch from $\mathcal{D}$
    \State $\psi \leftarrow \psi - \alpha_\psi {\nabla}_\psi J_{V_\psi}$
     \State $\theta \leftarrow \theta - \alpha_\theta {\nabla}_{\theta} J_{Q_\theta}$
     \State Compute $J_{\pi_\phi}$ via \eqref{eq:lsac-loss}
     \State $\phi \leftarrow \phi - \lambda_\pi {\nabla}_\phi J_{\pi_\phi}$
     \State $\bar{\psi} \leftarrow \tau \psi + (1 - \tau) \bar{\psi}$
  \EndFor
\EndWhile
\end{algorithmic}
\end{minipage}
\hfill
\begin{minipage}[t]{0.48\textwidth}
\raggedright
\textbf{Lyapunov Proximal Policy Optimization (LPPO)}
\vspace{0.5em}
\begin{algorithmic}[1]
\State Initialize policy $\pi_\phi$, RL value function $V_{\theta}$ Lyapunov function $L_\eta$ randomly
\State Initialize replay buffer $\mathcal{B} \leftarrow \emptyset$
\While{steps $ < K$}
  \State $\mathcal{B} \leftarrow \emptyset$
  \For{$t = 1$ to $N$}
    \State Sample $a_t \sim \pi_\phi(a_t|s_t)$
    \State Sample $s_{t+1} \sim P(s_{t+1}|s_t, a_t)$
    \State $\mathcal{B} \leftarrow \mathcal{B} \cup \{(s_t, a_t, r_t, s_{t+1})\}$
  \EndFor
  \State Sample mini-batches from $\mathcal{B}$
  \State Compute $J_{L_\eta}$ via \eqref{eq:lya-risk}
  \State $\eta \leftarrow \eta - \alpha_\eta \nabla_\eta J_{L_\eta}(\eta)$
  \For{each policy optimization step}
    \State Sample mini-batches from $\mathcal{B}$
    \State $\delta_t \leftarrow r_t + \gamma V_{\theta}(s_{t+1}) - V_{\theta}(s_t)$
    \State $\hat{A}(s_t,a_t) \leftarrow \delta_t + \gamma \delta_{t+1} + \dots$
    \State Compute $\hat{A}_\beta$ via \eqref{eq:adv} 
    \State Compute $J_{\pi_\phi}$ via \eqref{eq:lppo}
    \State $\phi \leftarrow \phi + \alpha_\phi \nabla_\phi J_{\pi_\phi}$
    \State $\theta \leftarrow \theta - \alpha_\theta \nabla_\theta J_{V_\theta}$
  \EndFor
  \State steps $\leftarrow$ steps $+ N$
\EndWhile
\end{algorithmic}
\end{minipage}

\vspace{0.5em}
\caption{The two proposed algorithms, LSAC (left) and LPPO (right). $J_{V_\psi}$, $J_{Q_\theta}$, $\bar{\psi}$ are defined in~\citep{sac}, and $J_{V_\theta}$ is defined in~\citep{ppo}.}
\label{fig:algorithms}
\end{figure*}



	

\section{Experimental Results}
\label{sec:result}

In this section, numerical experiments illustrate: (i) the application of the proposed off-policy Lyapunov SAC algorithm (LSAC) to an inverted pendulum; and (ii) how the off-policy Lyapunov function can be applied to a quadrotor, via LPPO, for which on-policy learning has been shown to be advantageous. 

\subsection{Inverted Pendulum}
The first experiment uses the standard Pendulum-v1 environment from Open AI Gym~\citep{gym}, without any modifications to the environment. Because the motor has insufficient torque to drive the pendulum directly to the upright position from all starting states, a swing up is sometimes necessary. The state is the position of the end of the pendulum, $x = \cos \theta$ and $y = \cos \theta$, and its angular velocity $\dot{\theta}$. The action is the torque $\tau$ applied by the motor at the joint.

\begin{figure*}[t]
    \centering
    \hfill
    \begin{subfigure}[b]{0.32\textwidth}
        \includegraphics[width=\textwidth]{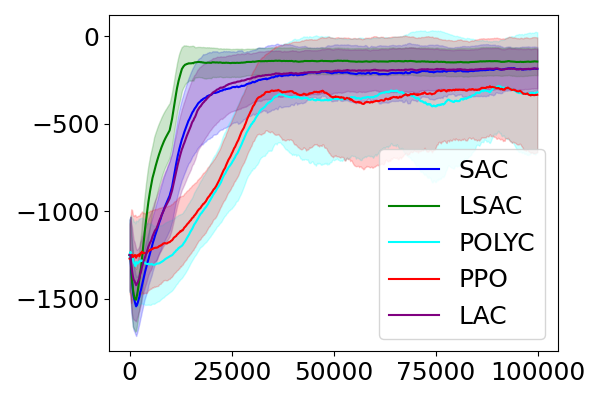}
        \caption{Training Rewards}
    \end{subfigure}
    \hfill
    \begin{subfigure}[b]{0.32\textwidth}
        \includegraphics[width=\textwidth]{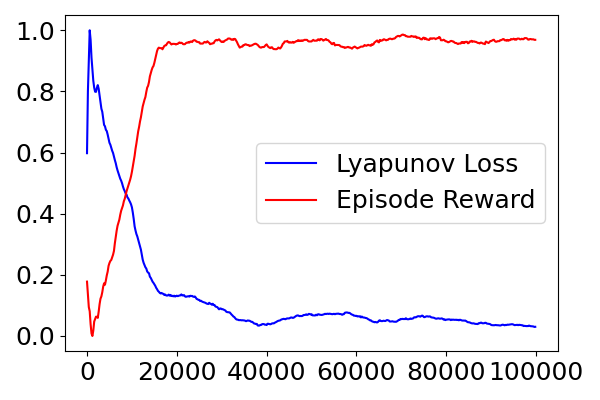}
        \caption{LSAC Reward vs Loss}
    \end{subfigure}
    \hfill
    \begin{subfigure}[b]{0.32\textwidth}
        \includegraphics[width=\textwidth]{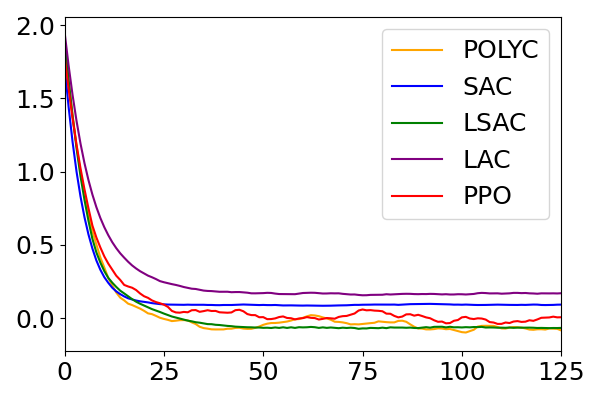}
        \caption{Sample Trajectory}
    \end{subfigure}
    
    \caption{\textcolor{black}{Pendulum-v1 Experiment Results: (a) the reward of different algorithms during training, as function of the number of episodes, and with the shaded region showing one standard deviation over the 10 random seeds; (b) the loss~\eqref{eq:lya-risk} and the reward during training (y axis is normalized); (c) a sample trajectory for each algorithm after training is complete.}}
    \label{fig:pend_rew}
\end{figure*}

Figure~\ref{fig:pend_rew} (a) depicts the training rewards of LSAC, SAC, LAC, POLYC and PPO for the Pendulum-v1 environment, for the first 100,000 training steps. Over the 10 random seeds, LSAC achieves the highest reward with the fewest steps to convergence, which indicates that LSAC is the most sample efficient. Figure~\ref{fig:pend_rew} (b) plots a sample trajectory after all algorithms have been trained. LSAC stabilizes the pendulum closest to the equilibrium $\theta = 0$ with minimal noise. POLYC also stabilizes it near the equilibrium but with more noise, while SAC and LAC stabilize it with minimal noise but further from the equilibrium. 

\begin{figure*}[t]
    \centering
    \hfill
    \begin{subfigure}[b]{0.32\textwidth}
        \includegraphics[width=\textwidth]{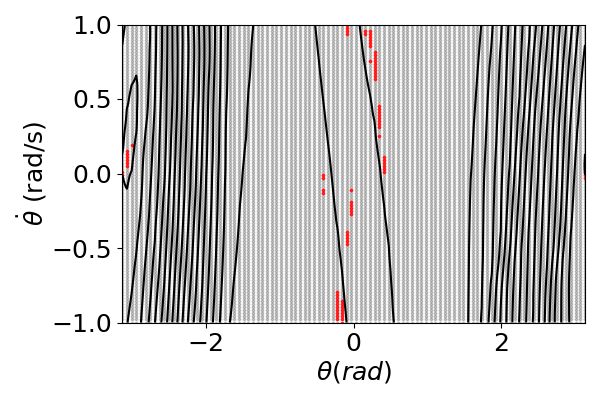}
        \caption{LSAC - 0.84\% Violations}
    \end{subfigure}
    \hfill
    \begin{subfigure}[b]{0.32\textwidth}
        \includegraphics[width=\textwidth]{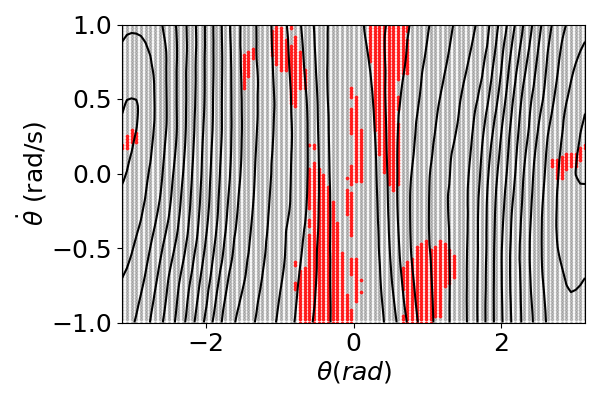}
        \caption{POLYC - 13\% Violations}
    \end{subfigure}
    \hfill
    \begin{subfigure}[b]{0.32\textwidth}
        \includegraphics[width=\textwidth]{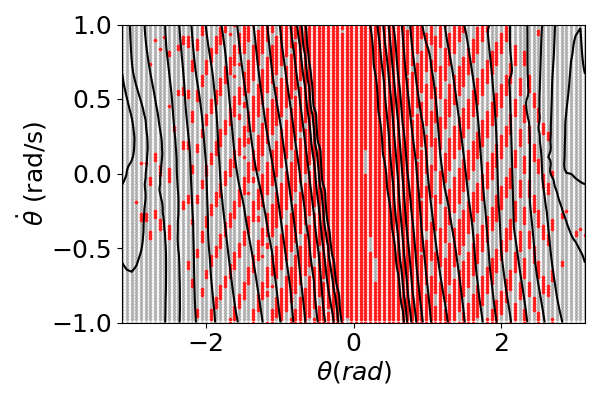}
        \caption{LAC - 51\% Violations}
    \end{subfigure}
    
    \caption{Level curves of the Lyapunov candidates learned by LSAC, POLYC and LAC. Grey dots represent pendulum states where the Lie derivative is negative. Red dots are pendulum states where the Lie derivative is positive.}
    \label{fig:pend_almost}
\end{figure*}

Figure~\ref{fig:pend_almost} shows, from left to right, the contours of the Lyapunov functions learned by the LSAC, POLYC and LAC. The red dots indicate violations of the Lyapunov decreasing condition along the simulated trajectories. The function learned by LSAC violates the decreasing condition the least, as illustrated by the minimal number of red dots in the left-most plot in Figure~\ref{fig:pend_almost}. The functions learned by POLYC and LAC violate the Lyapunov decreasing condition much more often, as seen in the larger number of red dots in the middle and right-most plots in Figure~\ref{fig:pend_almost}, respectively. If the Almost Lyapunov conditions~\citep{gao} were to be validated, LASC would have the largest region of attraction.


\subsection{Quadrotor}
Quadrotor control is a difficult problem for model-free RL. As shown in~\citep{almost-ly}, the two off-policy methods SAC and LAC struggle to produce any meaningful controller. Therefore, the numerical experiments in this section integrate the proposed off-policy Lyapunov function into the clean-RL implementation of PPO, which uses a normalized state and reward function for training~\citep{cleanrl}, to learn a trajectory tacking controller for a quadrotor simulated in the Mujoco physics simulator~\citep{mujoco}. As in~~\citep{traj}, the desired trajectory is generated by providing actions to the quadrotor and recording its state. The quadrotor then learns to track the desired trajectory guided by three algorithms: the proposed LPPO, the POLYC and the PPO algorithms.

The implementation extends~\citep{mujoco-quad} to track a trajectory. The 13-dimensional quadrotor state comprises the position error ($p_e \in \mathbb{R}^3$), the orientation error represented as a quaternion ($q_e \in \mathbb{R}^4$), the velocity error ($v_e \in \mathbb{R}^3$) and the angular velocity error ($\dot{\theta} \in \mathbb{R}^3$). The 4-dimensional controls are the applied thrust $F_z$ along the $z$ axis of the quadrotor's body frame measured in Newtons, and the angular velocity of the quadrotor along its $x$, $y$ and $z$ axes measured in rad/s. This choice of controls is justified (i) because motor thrusts map directly to the applied thrust and the body rates, and (ii) because body rates-based controls have better performance than motor thrust-based controls~\citep{quadrotor-thrust}.

\begin{figure}
    \centering
    \includegraphics[width=0.7\linewidth]{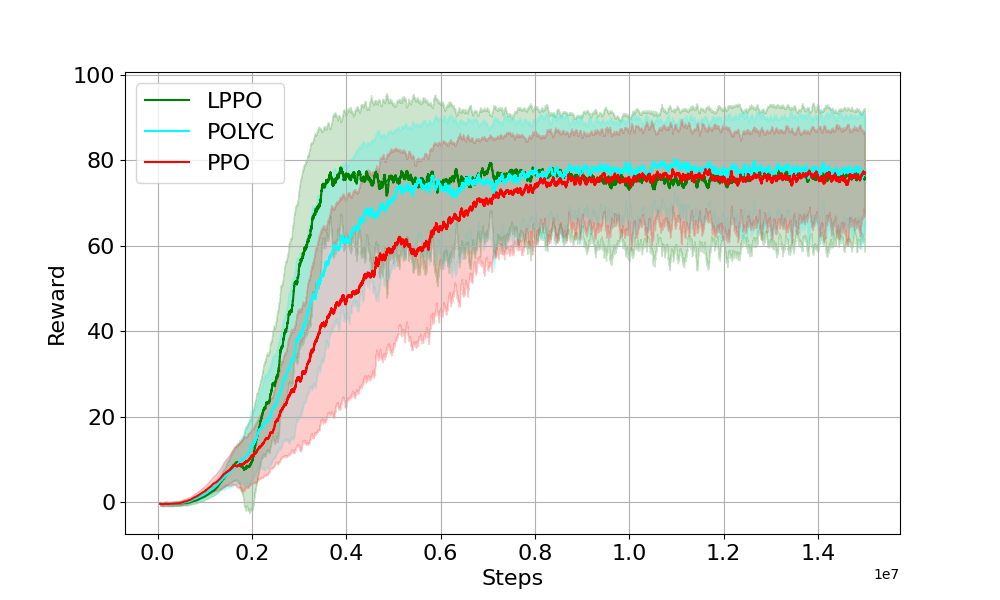}
    \caption{The mean training rewards for LPPO, POLYC, and PPO on the Mujoco Quadrotor environment, obtained from ten random seeds and plotted with a one standard deviation shaded region.}
    \label{fig:quad_rew}
\end{figure}

Since~\cite{almost-ly} has illustrated that SAC and LAC fail to learn any meaningful quadrotor control policy, this section compares only on-policy algorithms, namely the LPPO, POLYC and PPO algorithms. Figure~\ref{fig:quad_rew} shows the training rewards. LPPO and POLYC achieve a similar maximum reward while the PPO maximum reward is slightly lower. However, LPPO is more sample efficient as it converges faster than POLYC. 

\begin{figure*}[t]
    \centering
    \hfill
    \begin{subfigure}[b]{0.32\textwidth}
        \includegraphics[width=\textwidth]{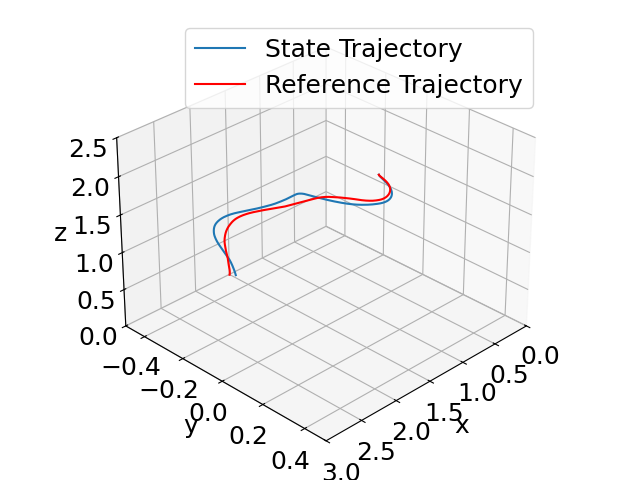}
        \caption{LPPO control.}
    \end{subfigure}
    \hfill
    \begin{subfigure}[b]{0.32\textwidth}
        \includegraphics[width=\textwidth]{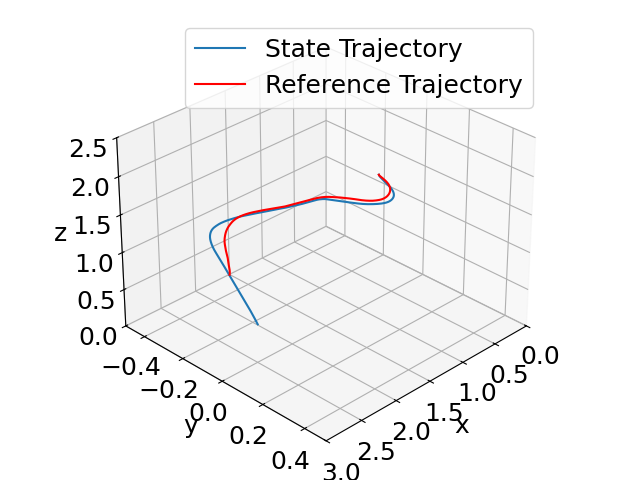}
        \caption{POLYC control.}
    \end{subfigure}
    \hfill
    \begin{subfigure}[b]{0.32\textwidth}
        \includegraphics[width=\textwidth]{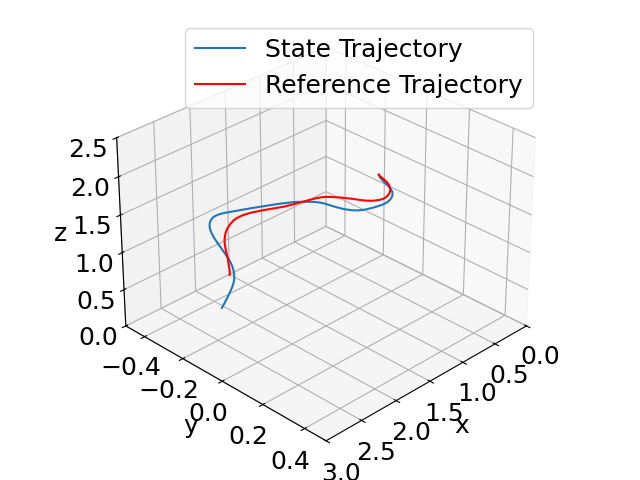}
        \caption{PPO control.}
    \end{subfigure}
    
    \caption{\textcolor{black}{Trajectory tracking for the quadrotor controlled by LPPO, POLYC, and PPO. The drone starts at the same starting point of $(x_0, y_0, z_0) \sim (1, 0, 2)$ for all three algorithms.}}
    \label{fig:quad_results}
\end{figure*}

Figure~\ref{fig:quad_results} plots sample trajectories after training is complete. LPPO tracks the reference trajectory most accurately. POLYC also tracks the reference trajectory accurately until the very end of the episode. PPO is also able to track the reference trajectory but with larger error compared both to LPPO and to POLYC.

\section{Conclusion}
\label{sec:conclusion}
This paper has proposed a method for self learning Lyapunov functions on off-policy data. Specifically, it has shown that a Lyapunov function can be effectively learned as the expectation over the actions under the current policy provided it depends both on the state and on the action. The paper has also illustrated how the proposed off-policy Lyapunov function can advise both off policy and on policy RL algorithms. Numerical experiments have demonstrated that the off-policy Lyapunov-based RL algorithms are more sample efficient and can achieve better performance on the Pendulum-v1 and Mujoco Quadrotor environments than existing RL algorithms.

\section{Limitations}
\label{sec:limitations}
While the experiments in Section~\ref{sec:result} show great success in simulated environments, the algorithms presented have yet to be tested in physical environments. A greater number of varied experiments would also aid in verifying the robustness of the proposed algorithms. Testing them in different simulated and physical environments is an important consideration for future work.

The proposed algorithms also include two additional hyperparameters; the minimum rate of decrease $\mu$ and the Lyapunov temperature $\beta$. The paper provides experimental results after hand tuning these hyperparameters. The inclusion of a hyperparameter sweep and an appropriate discussion is also an important direction for future work.

Because the proposed algorithms build upon an existing algorithm, the success of the underlying algorithm (i.e., SAC or PPO) is necessary for the success of the algorithms in this paper. 


\textcolor{black}{
This paper proposes a method to learn the Lyapunov function off-policy. Since the Lyapunov function is inherently dependent on the current controller, there is bias in the data collected from previous control policies. The paper proposes a method to address the bias but does not analyze the impact of the bias itself. Further work could compare the proposed method on off-policy and on-policy data, and could further reduce bias through importance sampling.
}

\textcolor{black}{
Lastly, the work presented shows promise in practice, but currently lacks theoretical support. Developing stability guarantees for the proposed algorithms is an important area for future work.}

\clearpage
\acknowledgments{The authors thank the reviewers for their constructive comments. They also acknowledge the financial support provided by the National Science and Engineering Research Council of Canada (DG34771)}


\addtolength{\textheight}{-3cm}   

\bibliography{references}  

\end{document}